7/14/20

# Using math in physics:
# 2. Estimation

*Edward F. Redish,*
University of Maryland - emeritus, College Park, MD

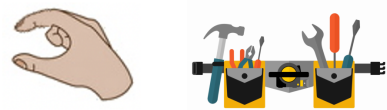

Learning to use math in science is a non-trivial task. It involves many different skills (not usually taught in a math class) that help blend physical knowledge with mathematical symbology. One of these is the idea of quantification — that physical quantities can be assigned specific numbers. A second is to develop an intuition for scale. One way to help students develop these skills is to teach *estimation*: the ability to consider a physical situation and put reasonable approximate numbers to it.[1]

The icon I use for estimation is the "Oh, about yay big" gesture shown at the top of the article. Everytime I do an estimation in the class or in the readings, this icon appears, reminding students that we are using this skill and allowing them to see how often we use it.

My personal story about estimation begins at the end of my senior year in college. My department had an end-of-program qualifier exam to determine (in part) whether we graduated with honors. I reviewed all my class notes from four years for this exam, hoping to do well. But… the exam's first question asked us to estimate five physical quantities quantitatively. The ones I remember are, "Estimate how much energy is stored in a firecracker" and "Estimate the speed of a molecule of air emitted from a small hole in an oven." I did very poorly on that set of questions. (I did OK on the rest of the exam.) Not only had I had no personal experience with firecrackers or ovens, I had no idea what they wanted me to do. I had never been assigned a problem that asked me to do anything like that in any of my 12 physics courses, and I don't believe that I had ever seen an instructor model an estimation in class.

In graduate school I learned about Fermi problems and, as a new Assistant Professor at the University of Maryland, I learned how much estimation was valued by professionals. I decided that if being able to do estimations was a skill valued in the physics community, we ought to teach it. I made it a practice to include estimation problems in all my classes.

When I began teaching algebra-based physics[2] and developing a special class for life science majors[3], I continued the practice. At first, I got a lot of resistance. Students complained, "This class is unfair. You get exam points for guessing." Nevertheless, I persisted and learned to be more explicit about why we were learning to do estimations and what value estimations could be to them.

In my later years teaching as a physics education researcher, thinking more deeply about the use of math in physics, I became convinced that estimation was not simply a fun "look what I can do" activity (that many students like), but was deeply tied to the sense of quantification of the physical world, the development of a quantitative intuition, and a critical element of the essential skill of model building.

I doubled down and developed my approach to estimation into an *e-game* (*epistemic-game*) — an explicit knowledge building strategy (not an algorithm). You can't actually do an estimation by turning a crank on an equation without thinking. You have to evaluate and make judgements of your own for the particular situation you are dealing with and with the particular knowledge you bring to the table.

To get students to take estimation seriously, I introduce them as a thread throughout the class, explicitly teaching it early and doing it in every context we are studying. Every homework contains an estimation problem as does every exam.

This is a much more successful approach than just asking them to do it occasionally on an exam! On my end-of-term-final-exam-extra-credit question ("Have you learned anything in this class that will be useful in five years?"), a significant fraction of my students now mention estimation as being of long-term value.

One of the most rewarding experiences in my class is when a student "gets" estimation and excitedly has to tell me a situation where they found it valuable. A few years ago, one of my life-science students cornered me after class to tell me her estimation story: There was going to be a rock concert on campus in a few weeks and, while the concert was free, there was limited space and you needed a ticket. My student and her friend went to the location where tickets were being distributed a few minutes before it opened and found a long line of students already waiting. Her friend said, "Oh, no! We'll never get tickets now. Let's go home." But my student had something to try. She said, "Wait", and did a quick estimation using blocking and a density argument. "There are 400 tickets available and there's only about 150 people on line. Let's get in line!" They got tickets, and she was so delighted at being able to estimate — a skill that she identified as having learned in my class — that she had to tell me about it.

Stories like this illustrate important points about estimation:

- Most students are perfectly capable of carrying out decent estimations.
- Most students have no idea that they can do this.
- When they learn to do it, they find it valuable not only in physics,
  but in their lives and in their other classes.
- Most students don't realize that this is a valuable scientific skill.

This last is particularly important. Many students bring the epistemological expectation into their physics class that physics is numerical and exact. This misconception leads to difficulties in their learning the role of modeling in physics and that science is always approximate, forever trying to improve its description of reality.

## Why is estimation important?

Some students will look at any request for an estimation and say, "I have no idea even how big that is." My most extreme example of that was an engineering student who came to my office hours to complain about estimations. She complained she was never taught anything's size. I asked, "Well, how big is a foot?" She said, "I don't know." I said, "Well, make a guess. How far up from the floor?" She put her hand about waist level. I said, "And how tall are you?" She thought for a moment, lowered her hand a bit, then lowered it again almost to her ankle, looked, and said with surprise, "Does the measure 'foot' have anything to do with a real foot?"

This is not meant in the least to put down a particular student's response. Rather, it is a terrific illustration of the "knowledge-in-pieces" or resources framework for analyzing student thinking.[4] We all know lots of things, but we often don't see the need to activate that knowledge or make connections between different bits that we know.

A basic epistemological lesson I want my students to learn is:

> *You have a lot of personal experiences that are relevant to almost everything you have to deal with quantitatively. Learning how to quantify and extend that personal experience is surprisingly powerful.*

Here are two reasons I think estimation is important.

1. Estimation helps students build a sense that the world can be quantified — and that their personal experience can be a valuable starting point in doing so.
2. Estimation helps students develop an intuitive sense of scale.

These first two items are basic steps in helping students build an understanding of how to think about the world with mathematics. Since our first step was identifying the symbols in physics as *measurements rather than numbers*,[5] getting a sense of scale helps our students to tie numbers (with units) to measurements; and it helps them develop a sense of what our measurements mean in terms of physical size (of whatever kind of quantity we are discussing).

We want our students to build quantitative physical intuition — the sense that they have quantitative knowledge about the world from their everyday experience, the confidence to call on it even inside a formal physics problem in a classroom setting, and the skill to use that knowledge in an effective manner.

3. Estimation is essential in deciding how to create appropriate physical models of phenomena.
4. Intuitions associated with estimation are essential in understanding how to approach almost any problem solving situation in physics.

One reason why we care about estimation and a quantitative sense of scale is that they are crucial components in creating mathematical ways of thinking about the world: building models. When we build a model to describe a physical phenomenon, we have to decide what phenomena we're going to pay attention to in our model and what we're going to ignore (at least at first). Doing that involves an intuition about quantitative scales, even if we're only saying, "That is not going to matter if we only care about this." Doing explicit estimation problems is a good way to develop that intuition. I'll discuss this application in detail in the paper in this series on Toy Models.[6]

## Which of our students need to learn estimation skills?

Estimation skills are widely recognized as a core skill of a professional physicist. But I would argue that the development of the "estimation perspective" — the view that you can get insights into physical phenomena from quick, crude estimates — is one of the valuable tools a physics class can offer our students from other disciplines. This is particularly true for life science students since, in my experience, many tend not to think about quantification at all when they learn about biological mechanisms. Many biology textbooks fail to include a scale on their many photos of microscopic structures in biology implicitly sending an unfortunate unintended message that size, even approximate size, is not really relevant.

## Introducing estimation

Since most students have little experience with estimation, I introduce it early in the class, give some examples, and specify some rules of the game; for example

- Don't look up any data in books, online, or get it from friends (unless the problem states otherwise). Learn to develop your own numerical estimation skills.

With the current ubiquitous availability of search engines, this is a hard principle to enforce. But if they assume they can look anything up they will ever need, they will never develop a quantitative intuition or sense of scale for themselves.

- Do not use your calculator. Do all calculations to one or two significant figure accuracy.
- Use powers of 10 notation.

These are also hard rules for them to accept. Many students today don't trust themselves to do even simple one digit math without a calculator. Despite all having been exposed to powers of 10 notation, many are not really comfortable with it. They can all do both one digit math and powers of ten and get comfortable with them, but they have to understand why we want them to do it — and they have to feel that if they don't, it will not be ignored and they will lose credit.[7] Doing this approximate math helps them "own" numbers and begin to feel more comfortable proposing them, manipulating them, and interpreting them.

- Explicitly state your underlying assumptions and be sure they relate to some plausible personal experience whose values you can trust.

A big part of learning to do an estimation problem is convincing your listener/reader that the numbers you are "making up" are ones that you could plausibly know. In my experience, many students want to answer this by "I remember this from high school." In almost every case, that "remembering" produces a number that is wildly out of line. On the other hand, the student who said, "I once worked in a bakery and had to carry 50 pound sacks of flour, so I know how heavy that is" was given full credit.

- Specify the physical principles you are using so we can see how you get from your starting assumptions to your results.

This is another epistemological shift we need the students to make: It's not only the answers that matter, but how you got there. Explain your reasoning! As the class moves on, I switch from asking them to do everyday estimations (How many blossoms are there on this photo of a cherry tree?) to ones that involve reasoning with physics principles (Estimate the spring constant of a trampoline if the picture shows a six-year old at the top of her jump).

Here's an example of what I mean by "convincing your listener/reader".

## A sample problem

> How many blades of grass are there in a typical lawn in the suburbs in June?

You might decide you could estimate the size of the lawn, but you would need the density of the grass -- the number of blades per square meter. If you said, "Let's assume that there are a million blades of grass per square meter" you would receive no credit. If you said, "When I lie down on a grassy lawn, I can see the grass. Knowing that the last joint of my thumb is about 1 inch long, I can easily imagine the grass against it. I can then see about 10 blades of grass against half that thumb joint, or 10 per cm. This makes 100 (= 10x10) per square cm, or $10^2(10^2)^2 = 10^6$ or one million blades per square meter." That would receive full credit because it connects to something that the listener can imagine that the writer plausibly knows.

## Don't memorize
## (but know some scales).

Although I stress that students should be building numbers based on personal experience, there are some scales they should just know from previous classes. But since those scales are often not taught in those classes, I give them a list of some numbers I expect them to have available.

- How big is an inch, foot, mile, kilometer
- How long is a minute, hour, year
- The density of water
- Their own height and weight
- Various temperatures (freezing ice, boiling water) in Fahrenheit, Celsius, and Kelvin

In a group activity early in the class, I pass around meter sticks and ask them to measure and write down the length of the first joint of their thumb, their handspan, their forearm (cubit), and their height in centimeters to two sig figs (only) for use later on homework and exams. This helps them realize that I am serious about their constructing numbers on exams from their personal experience and gives them personal built-in measuring sticks. (See the supplementary EPAPS materials.)

There are also some (very approximate) scales I ask my students to learn as "foothold" scales that permit them to do quantitative estimations at scales they don't experience directly.

Since I often give problems that have some environmental relevance, I want them to know these approximate numbers:

- Radius of the earth ( $\sim 2/\pi \times 10^7$ meters)[8]
- Distance across the USA (~ 3000 miles)
- Number of people in the USA (~ 330 million)
- Number of people on earth (~ 8 billion)

For my life science students, I ask them to know a few additional scales:

- Size of a typical animal cell (~10-20 microns)
- Size of a bacterium or chloroplast (~1 micron)
- Size of a medium-sized virus (~0.1 micron)
- Thickness of a cell membrane (~ 0.005 microns = 5 nm)
- Size of a protein molecule (varies, but ~5 nm is OK)
- Diameter of an atom (~ 0.1 nm)

If you develop estimation problems relevant to your own teaching goals, you might want to have a small set of numbers you expect your students to know.

## Using estimation in class

For students to take estimation seriously, we have to both teach them how to do it, have them do it, and evaluate them on their performance. I introduce estimation early in the class and give explicit instructions for how to approach it, both on homework and on exam problems. (See the advice for doing exam estimations in the groupwork problem in the supplementary materials.)

*Grading estimations*: It's not enough to tell students how to do an estimation. Since so many students come into the class with the epistemological misconception that physics is "answer-focused" rather than "reasoning-focused", it's important to give most of the credit for a homework or exam estimation for the reasoning rather than for the answers.

Here's how I set up grading rubrics for grading estimations. How many points you specifically assign depends on how many points are assigned to the problem. (I usually give 15 points for an estimation question on a 100-point hour exam.) I use an approximately equal division of points for each of the following (though the specific division depends on the particular problem):

- Stating the principles being used to set up the calculation, either physical or geometrical.
- Creating reasonable numbers needed for the estimation.
- Explaining how those numbers were generated from experiences they could trust or things they could be expected to know.
- Putting the appropriate numbers in the right slots in the equation.
- Carrying out the calculation correctly (including unit conversion).
- The answer.

I insist that if they use equations they be given symbolically rather than have numbers put in right away. I subtract 1 point if they don't keep units all through the calculation. (This is particularly important since I like estimation problems that cross scales. I'll often give them some numbers but in units that require conversion.) I subtract 1 point if they EVER keep more than 3 significant figures in their calculation and 1 point if they EVER use long strings of 0's in a number rather than using powers of 10 notation.

And the acceptable range for the answer should be quite wide — typically a factor of two or more.

Here are some examples of how I bring the estimation game as a thread into my class. (Estimation can be productively combined with dimensional analysis and scaling. See the paper on functional dependence for specific examples.[9])

### A quiz question (early in the class)

| You know that 1 cubic centimeter of water has a mass of 1 gram. What's the mass of one cubic meter of water? | | | |
|---|---|---|---|
| a. | 10 g | e. | 1 kg |
| b. | $10^2$ g | f. | 10 kg |
| c. | $10^4$ g | g. | 100 kg |
| d. | $10^6$ g | h. | 1000 kg |

This one depends on dimensional analysis and the functional dependence of volume on length but not yet creating new numbers skills. But it helps them see what it means to get a sense of scale. Many students are surprised to see what one cubic meter of water weighs since they can easily imagine that volume of water and expect that they could lift it.

### A Peer Instruction problem (early in the class)

| Estimate the thickness of a page in a typical textbook. | | | |
|---|---|---|---|
| A. | $10^{-1}$ m | E. | $10^{-5}$ m |
| B. | $10^{-2}$ m | F. | $10^{-6}$ m |
| C. | $10^{-3}$ m | G. | $10^{-7}$ m |
| D. | $10^{-4}$ m | H. | $10^{-8}$ m |

What assumption(s) did you have to make to come up with your estimate?

This is a great one for a Peer Instruction-style class discussion.[10] I typically ask them to do this first by themselves and show them the distribution of answers (which tends to be pretty wide). I then let them discuss it in groups of 3-4 for a few minutes and ask for their answers again. The spread narrows dramatically. This allows me to make the point of the value of discussing your answers with peers.

### An exam problem

Flu season is approaching! Oscillococcinum is a homeopathic medicine that is advertised as relieving flu symptoms. It is produced by starting with duck liver and heart, and diluting them to "200C". This means that they are 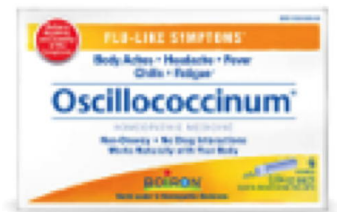 diluted to 1 part per hundred, and then this is repeated 200 times. (Thus, after two dilutions, the ratio is 1 in $10^4$; after three, the ratio is 1 in $10^6$, and so on all the way to 200.) Estimate how many molecules of the original duck organs are in one Oscillococcinum pill.[11]

## Instructional resources

Many of the ideas for this series of paper were developed in the context of studying physics learning in a class for life-science majors. Problems and activities using estimation are offered in the supplementary materials to this paper. In addition to the materials presented in the supplementary materials to this paper, there is an extensive collection of readings and activities from this project on estimations available at the *Living Physics Portal*,[12] search "Making Meaning with Mathematics: Estimation."

There are a lot of resources available for working with estimations. My approach is mostly to help students develop some quick and dirty intuitions. Sanjoy Mahajan has a series of very valuable articles in the *American Journal of Physics*[13] and a book[14] on the topic that provide specific tools and deeper insights. Highly recommended!

## Acknowledgements

I would like to thank the members of the UMd PERG over the last two decades for discussion on these issues. I also thank Ben Dreyfus and Arpita Upadhyaya for creating some of the estimation problems given here. The work has been supported in part by a grant from the Howard Hughes Medical Institute and NSF grants 1504366 and 1624478.